\documentclass[cits]{PoS}

\title{The b quark mass from lattice nonrelativistic QCD}

\ShortTitle{The b quark mass from lattice nonrelativistic QCD}

\author{Alistair Hart$^a$, Georg M.~von Hippel$^b$, R. R.~Horgan$^c$, 
Andrew Lee$^c$, \speaker{Christopher J. Monahan}$^c$\\
\llap{$^a$}SUPA, School of Physics and Astronomy, University of
Edinburgh, Edinburgh EH9 3JZ, 
U.K.\footnote{Current address: Cray Exascale Research Initiative, JCMB, King's Buildings, Edinburgh EH9 3JZ, U.K.}\\
\llap{$^b$}Institut f\"ur Kernphysik, Universit\"at Mainz, Becherweg 45, 55099 Mainz, Germany\\
\llap{$^c$}DAMTP, University of Cambridge, Wilberforce Road, Cambridge CB3 0WA, U.K.\\
Email: \email{C.Monahan@damtp.cam.ac.uk}, \email{a.hart@ed.ac.uk},
\email{hippel@kph.uni-mainz.de},
 \email{R.R.Horgan@damtp.cam.ac.uk}, \email{A.Lee@damtp.cam.ac.uk}}

\abstract{We present the first two-loop calculation of the heavy quark 
energy shift in lattice nonrelativistic QCD (NRQCD). This calculation 
allow us to extract a preliminary prediction of $m_b(m_b, \; n_f = 5)
= 4.25(12)$ GeV for the mass of the b quark from lattice NRQCD simulations
performed with a lattice of spacing $a=0.12$fm. 
Our result is an improvement on a previous determination of the b
quark mass from unquenched lattice NRQCD simulations, which was 
limited by the use of one-loop expressions for the energy shift. Our 
value is in good agreement with recent results of $m_b(m_b) =
4.163(16)$ GeV from QCD sum rules and $m_b(m_b, \; n_f = 5) =
4.165(23)$ GeV from realistic lattice simulations using
highly-improved staggered quarks. We employ a mixed strategy to 
simplify our calculation. Ghost, gluon and counterterm contributions 
to the energy shift and mass renormalisation are extracted from
quenched high-beta simulations whilst fermionic contributions 
are calculated using automated lattice perturbation theory. 
Our results demonstrate the effectiveness of such a strategy.}

\FullConference{The XXVIII International Symposium on Lattice Field Theory, Lattice2010\\
		June 14-19, 2010\\
		Villasimius, Italy}

\begin{document}

\section{Introduction}

The precise theoretical and experimental determination of quark masses
is an important component of high-precision tests of the Standard
Model of particle physics. One current focus for tests of the Standard 
Model is the unitarity of the Cabibbo-Kobayashi-Maskawa (CKM) matrix, 
which describes flavour-changing quark transitions. Quark masses serve 
as an input into the tests of CKM matrix unitarity; the mass of the b 
quark is used in the extraction of the CKM matrix element 
$\left|V_{ub}\right|$ from inclusive semileptonic decays of B mesons \cite{barberio06}.

Recent high-precision calculations of the b quark mass using realistic 
lattice QCD simulations \cite{mcneile10} and perturbative QCD combined 
with experimental results \cite{chetyrkin09} are in good agreement, 
obtaining values of $m_b(m_b, \; n_f = 5) = 4.165(23)$
GeV and $m_b(m_b) = 4.163(16)$ GeV respectively.
The lattice result was obtained using the highly improved staggered quark 
(HISQ) action for the valence quarks, with the improved relativistic 
Asqtad action used for the 
sea quarks. HISQ is a highly corrected version of the 
standard staggered quark action that retains a chiral symmetry on 
the lattice \cite{follana07}. Most current lattice studies of b 
quarks use an effective field theory, such as nonrelativistic QCD 
(NRQCD), for the valence heavy quarks. Simulating both valence and 
sea quarks with relativistic actions
allows a much greater precision, but is only now becoming possible 
with the advent of finer lattices and highly improved actions. 
However, even on the very finest lattices with HISQ heavy quarks 
an extrapolation to the heavy quark mass is still required \cite{mcneile10}. 

Our calculation improves on a previous determination of $m_b(m_b) =
4.4(3)$ GeV from unquenched lattice QCD simulations using NRQCD 
valence b quarks \cite{gray05}. The dominant error in that calculation 
arose from the use of one-loop perturbation theory in the matching 
between lattice quantities and the continuum result. By introducing 
a mixed strategy incorporating high-beta quenched simulations and 
automated lattice perturbation theory, we perform the first ever such 
two-loop calculation in NRQCD. This serves a two-fold purpose. 
Firstly our calculation demonstrates the effectiveness of employing 
an efficient mixed perturbation theory/high-beta simulation method 
for higher order perturbative quantities. Secondly our result allows
us to obtain a more precise prediction for the b quark mass from 
lattice NRQCD simulations.

\subsection{Heavy quarks on the lattice}

Currently available lattices are too coarse to directly simulate 
b quarks, because the Compton wavelength of the b quark is smaller 
than the lattice spacing. One common approach to solving this 
problem is to introduce a nonrelativistic effective action, NRQCD, 
for which the discretization errors are under control and which can be 
systematically improved by including extra operators.

NRQCD is constructed by integrating out dynamics at the scale of 
the heavy quark mass and then using the Foldy-Wouthuysen-Tani 
transformation to write the action as an expansion in the inverse 
heavy quark mass \cite{lepage92}. We use an NRQCD action correct 
to ${\cal O}(1/m^2,v^4)$, where $v$ is the relative internal 
velocity of the bound-state heavy quarks. A detailed derivation 
of the action we use is given in \cite{horgan09}. The lattice NRQCD action 
can be written
\begin{equation}\label{eq:snrqcd}
S_{\mathrm{nrqcd}} = \sum_{\mathbf{x}, \tau} \psi^+(\mathbf{x},\tau)
\left[\psi(\mathbf{x},\tau) - K(\tau)\psi(\mathbf{x},\tau-1)\right],
\end{equation}
with
\begin{equation}
K(\tau) = \left(1 - \frac{\delta H}{2}\right)
\left(1 - \frac{H_0}{2n}\right)^nU_4^{\dagger}\left(1 -
  \frac{H_0}{2n}\right)^n
\left(1 - \frac{\delta H}{2}\right).
\end{equation}
Here the leading nonrelativistic kinetic energy is 
$H_0 = -\Delta^{(2)}/2M$. The correction term $\delta H$ 
contains higher order terms in the $1/M$ expansion: the 
improved chromoelectric and chromomagnetic interactions, 
the leading relativistic kinetic energy correction and 
discretization error corrections. The integer $n$ is 
introduced as a stability parameter.

\section{Calculating the b quark mass}

Quark confinement ensures that quark masses are not 
physically measurable quantities, so the notion of 
quark mass is a theoretical construction. A wide range 
of quark mass definitions exist, often tailored to 
exploit the physics of each particular process. One common 
choice of quark mass is the pole mass, defined as the 
pole in the renormalized heavy quark propagator. 
However, the pole mass is a purely perturbative concept 
and suffers from infrared renormalon ambiguities 
 \cite{bodwin99,bigi94b,beneke94}.  To avoid these ambiguities, 
experimental results are usually quoted in the modified 
Minimal Subtraction ($\mathrm{\overline{MS}}$) scheme, 
which is renormalon ambiguity free. Lattice calculations 
use the renormalon-free bare lattice mass. These different 
quark mass definitions must be matched to enable meaningful 
comparison. We match bare lattice quantities to those in the
$\mathrm{\overline{MS}}$ scheme using the pole mass as an 
intermediate step. Any renormalon ambiguities cancel 
in the full matching procedure between the lattice quantities 
and the $\mathrm{\overline{MS}}$ mass.
We extract the $\mathrm{\overline{MS}}$ mass from lattice 
simulation data in a two-stage process. We first relate 
lattice quantities to the pole mass and then match the 
pole mass to the $\mathrm{\overline{MS}}$ mass evaluated 
at a scale equal to the b quark mass.

\subsection{Extracting the pole mass}

We determine the pole mass using two independent methods. 
The first method relates the pole mass, $M_b^{\mathrm{pole}}$,
 to the experimental $\Upsilon$ mass, 
$M_{\Upsilon}^{\mathrm{expt}} = 9.46030(26)$ GeV \cite{pdg10}, 
using the heavy quark energy shift, $E_0$:
\begin{equation}\label{eq:e0}
2M_b^{\mathrm{pole}} = M_{\Upsilon}^{\mathrm{expt}} - (E^{\mathrm{sim}}(0) - 2E_0).
\end{equation}
Here $E^{\mathrm{sim}}(0)$ is the energy of the $\Upsilon$ 
meson at zero momentum, extracted from lattice NRQCD simulations. 
The quantity $(E^{\mathrm{sim}}(0) - 2E_0)$ corresponds to the 
`binding energy' of the meson in NRQCD. We use a value of 
$E^{\mathrm{sim}} = 0.515(3)$ GeV, obtained from a lattice 
NRQCD simulation run by the HPQCD collaboration on a
 ``coarse'' MILC ensemble, with lattice spacing $a = 1.647(3)$ 
GeV$^{-1}$ \cite{davies10a}. For further details of the 
configuration ensemble see \cite{bazavov10,davies10b}.

The second method directly matches the pole mass to the
 bare lattice mass in physical units, $M_b^{\mathrm{latt}}(a)$, 
via the heavy quark mass renormalisation, $Z_M^{\mathrm{latt}}$,
\begin{equation}\label{eq:zmlatt}
M_{\mathrm{pole}} = Z_M^{\mathrm{latt}}(\mu a, M_b^{\mathrm{latt}}(a)) M_b^{\mathrm{latt}}(a).
\end{equation}

\begin{figure}
\begin{centering}
\includegraphics[width=0.9\textwidth,height=0.4\textwidth]{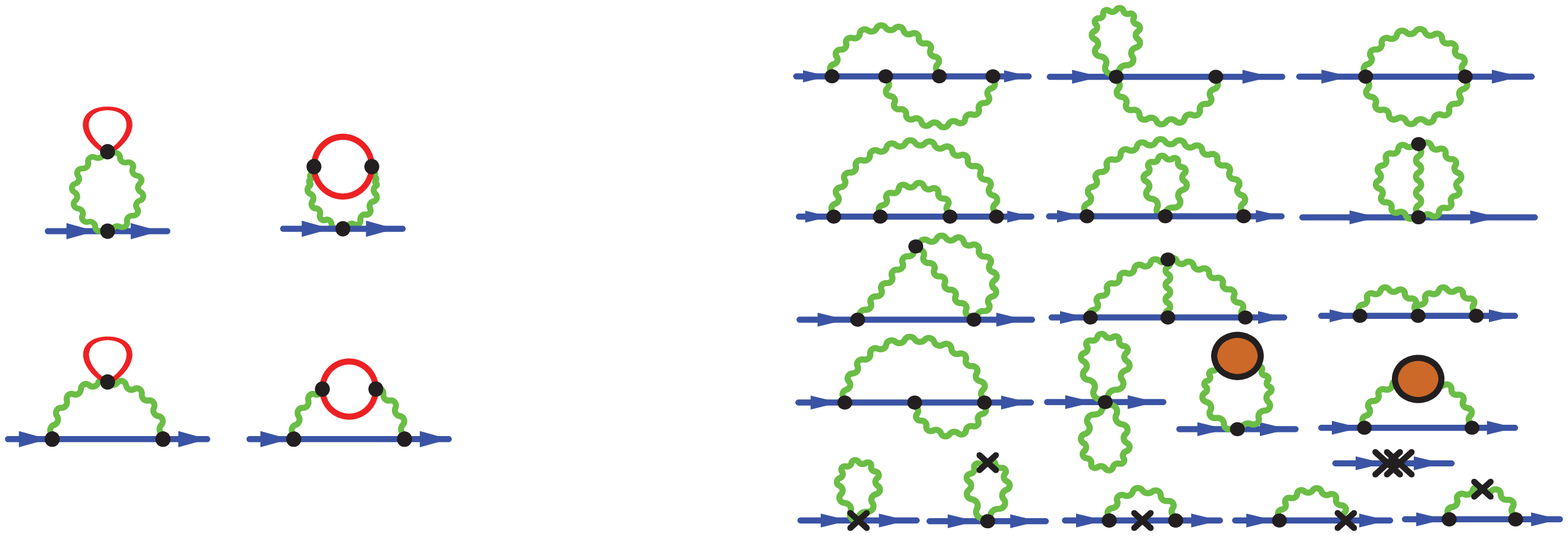}
\caption{Contributions to $E_0$ and $Z_M^{\mathrm{latt}}$. The four 
fermionic contributions calculated using automated lattice
perturbation theory are shown on the left. The diagrams on 
the right are extracted from high-$\beta$ simulation. Blue 
lines are heavy quarks, green are gluons and red are sea quarks. 
Large brown blobs represent the 5 gluon self energy diagrams and 
crosses are counterterms. Feynman diagrams reproduced from \cite{mason05}.}
\label{fig1}
\end{centering}
\end{figure}

We employ a mixed strategy to calculate $E_0$ and
$Z_M^{\mathrm{latt}}$ perturbatively. The fermionic contributions to $E_0$,
 shown on the left-hand side of Figure \ref{fig1}, are 
calculated using two-loop automated lattice perturbation theory. 
All other contributions, shown on the right-hand side of Figure
 \ref{fig1}, are extracted from high-beta quenched simulations.

Results were obtained using the NRQCD action of Equation 
\ref{eq:snrqcd} for the heavy valence quark, HISQ light 
quarks and the L\"uscher-Weisz action for the gluons 
\cite{hao07,luscher86}. We used a heavy quark mass in lattice 
units of $Ma = 2.8$, with a stability parameter of $n=2$.

\subsubsection{Automated lattice perturbation theory}

Feynman rules for the NRQCD and HISQ actions are too complicated
 to be viably derived by hand and the resulting Feynman integrals 
can only be evaluated numerically. We therefore use automated 
lattice perturbation theory, employing \verb+HiPPy+ to derive the 
Feynman rules and \verb+HPsrc+ to evaluate the four diagrams \cite{hart09,hippel10}.
To control the highly-peaked IR behaviour of the Feynman integrands, 
we introduce a gluon mass. Although in general a non-zero gluon mass 
cannot be used in calculations beyond one-loop, this issue concerns 
only diagrams containing ghost-gluon vertices. In our calculation, 
these diagrams are handled by the high-beta simulation, allowing us 
to use a gluon mass for the fermionic contributions. The light quark
 diagrams in Figure \ref{fig1} were calculated using five different 
light quark masses and extrapolated to zero light quark mass. We verified that the
appropriate Ward identity for the 1-loop gluon self-energy was satisfied.

\subsubsection{High-beta simulations}

We perform quenched simulations on $L^3 \times T$ lattices with 
temporal extent $T=3L$, for $L=3$ to $L=10$ and twisted boundary 
conditions to reduce finite size effects and tunnelling between 
QCD vacua \cite{luscher86}. We generate ensembles of configurations for 
17 values of $\beta$ from
$\beta=9$ to $\beta=120$.  Since the Green function is not gauge-invariant, we fix to 
Coulomb gauge using a conjugate gradient method. To extract the energy shift and mass renormalisation, 
we use an exponential fit to the heavy quark Green function parametrized as
\begin{equation}
G({\bf p},t) = Z_\psi \exp\left(-\left[E_0+\frac{p^2}{2Z_M^{\mathrm{latt}}M_0} + \ldots \right]t\right),
\end{equation}
where the ellipsis stands for higher order terms that are included in the fits.

All operators in the NRQCD action are expressed in terms of gauge-covariant Wilson paths 
generated using PYTHON, which greatly enables flexibility and reduces programming errors. 
The heavy quark source is classified in the flavour-smell basis appropriate to the twisted 
boundary conditions. We implement the boundary conditions using a gauge-twist mask
whenever a path in an operator crosses any spatial boundary. By
applying an extra $U(1)$ phase in the mask, we can assign an arbitrarily small momentum
to the source, enabling both $E_0$ and $Z_M^{\mathrm{latt}}$ to be reliably 
extracted as a function $(\beta,L)$. 

We convert $\beta$ to $\alpha_V$ and
perform a joint fit to extract the 1- and 2-loop coefficients in the $L \to \infty$ limit.
Simulations were run including tadpole improvement, which significantly reduces the 
magnitude of both 1- and 2-loop coefficients. Results for $Z_M^{\mathrm{latt}}$ are good but this work 
is still in progress and we concentrate here on those for $E_0$. For even $L$, in Table \ref{E0} we compare
the tadpole-improved 1-loop coefficient from an unconstrained fit to the simulation data for 
$E_0$ with the exact result from automated perturbation theory. To extract the 2-loop 
coefficient we constrained the 1-loop coefficient to be the exact value, but Table \ref{E0} shows 
that the simulation reliably reproduces the 1-loop results. The number of independent configurations 
for each $(\beta,L)$ was about 300, which we can easily increase by 10-fold or more, allowing for 
much more accurate results at the next stage.

\begin{table}[hbt]
\center{
\begin{tabular}{|c||c|c|c|c|c|}\hline
$L$&4&6&8&10&$\infty$\\\hline
$E_0^{\mbox{sim.}}$&0.5295(16)&0.5988(16)&0.6369(12)&0.6560(11)&0.7380(63)\\\hline
$E_0^{\mbox{th.}}$&0.5312&0.6020&0.6362&0.6565&0.7348(3)\\\hline
\end{tabular}

\caption[]{\label{E0}
Comparison of an unconstrained fit from simulation for the perturbative 1-loop coefficient 
with the automated perturbative calculation. There is no error on the theory calculation as
it was done by mode summation. The error on the theory extrapolation to $L=\infty$ is estimated from
a fit.}
}
\end{table}

\subsection{Matching the pole mass to the $\mathrm{\overline{MS}}$ mass}

Although the pole mass is plagued by renormalon ambiguities, 
these ambiguities cancel when lattice quantities are related 
to the $\mathrm{\overline{MS}}$ mass. This renormalon cancellation 
is evident in the direct matching of the bare lattice mass to the 
$\mathrm{\overline{MS}}$ mass,
\begin{equation}
M^{\overline{MS}}(\mu) = Z_M^{\mathrm{latt}}(\mu a,
M_b^{\mathrm{latt}}(a))
Z_{\mathrm{cont}}^{-1} (\mu, M_{\mathrm{pole}})M_b^{\mathrm{latt}}(a),
\end{equation}
as both $M^{\overline{MS}}$ and $M_b^{\mathrm{latt}}$ are 
renormalon-free. The continuum matching parameter,
$Z_M^{\mathrm{cont}}$, 
relates the pole mass to the $\mathrm{\overline{MS}}$ mass and has 
been determined to ${\cal O}(\alpha_s^3)$ \cite{melnikov99}.

To see that renormalon ambiguities also cancel in when determining 
the pole mass from the energy shift, we equate Equations \ref{eq:e0} 
and \ref{eq:zmlatt} and rearrange them to obtain
\begin{equation}
2(Z_M^{\mathrm{latt}}M_b^{\mathrm{latt}}(a) - E_0) =
M_{\Upsilon}^{\mathrm{expt}} - 
E^{\mathrm{sim}}(0).
\end{equation}
The two quantities on the right hand side of the equation are
renormalon 
ambiguity free: $M_{\Upsilon}^{\mathrm{expt}}$ is a physical quantity 
and $E^{\mathrm{sim}}(0)$ is determined nonperturbatively from lattice 
simulations. Any renormalon ambiguities in the two power series, 
$Z_M^{\mathrm{latt}}$ and $E_0$, on the left-hand side of the equation 
must therefore cancel.

\section{Results}

For the fermionic and quenched contributions to the two-loop 
heavy quark energy shift we find
\begin{equation}
E_0 = 0.7348(3)\alpha_V(q^{\star}/a) + (1.37(6) - 0.0023(1)n_f)
\alpha_V^2(q^{\star}/a) + {\cal O}\left(\alpha_V^3\right).
\end{equation}
We express our result in the $V$-scheme at a scale $q^{\star}/a = 3.33$, 
a value determined using the BLM procedure in \cite{wong06}. 
The uncertainties quoted for the one-loop coefficient and the 
quenched contribution to the two-loop coefficient arise from 
the multi-polynomial fit. For the fermionic contribution to 
the two-loop coefficient, the quoted uncertainty is the 
statistical error in the numerical evaluation of the Feynman 
diagrams. We estimated the coefficient of the 
${\cal O}\left(\alpha^3_s\right)$ term from the quenched 
simulation fits as $\sim 1.0(5)$.

Inserting this result for the heavy quark energy shift 
into Equation \ref{eq:e0} leads to our first preliminary 
determination of the b quark mass: 
\begin{equation}
M^{\overline{MS}}\left(M^{\overline{MS}}\right) = 4.25(12) \; \mathrm{GeV}.
\end{equation}
The error is an estimate of ${\cal O}(\alpha^3_s)$ contributions, 
which dominate the uncertainty in our result. Uncertainties 
arising from systematic and statistical errors in the lattice
 results, $E^{\mathrm{sim}}(0)$ and $E_0$, are $\ll 1\%$. We are unable to 
estimate the systematic error due to $O(a^2)$ artifacts as we have not yet 
finished the calculation for smaller values of $a$; this work is in 
progress and entails working with different values of $Ma$ in NRQCD. 
It should be noted that we used a value of $E^{\mathrm{sim}}(0)$ that was
 generated from lattice NRQCD simulations using the action of 
Equation \ref{eq:snrqcd}, but with $n=4$. From 1-loop calculations we estimate 
the errors associated with this mismatch to be much smaller than the dominant
${\cal O}(\alpha^3_s)$ error. However, this discrepancy will be corrected in future work.

\section{Conclusion}

We have calculated the two-loop heavy quark energy shift 
in highly-improved NRQCD using a mixed approach combining 
quenched high-beta simulations with lattice perturbation theory. 
This is the first determination of any heavy quark parameter 
beyond first-order perturbation theory in NRQCD, and demonstrates that we are able 
to extract a more precise prediction of the b quark mass from 
lattice NRQCD simulations than has been previously achieved. 
Work is currently underway to complete our calculation of the
mass renormalisation, $Z_M^{\mathrm{latt}}$, and to extend our results to incorporate 
different heavy quark masses to enable us to effectively estimate ${\cal O}(a^2)$ 
uncertainties. We also plan to increase the size 
of the ensembles  used in the high-beta analysis by a significant factor. We expect these 
developments will improve further the precision of our result 
for the b quark mass.

\section*{Acknowledgements}

We would like to thank Christine Davies and Iain Kendall for 
providing HPQCD simulation data.

We thank the DEISA Consortium (www.deisa.eu), funded through the 
EUFP7 project RI-222919, for support within the DEISA Extreme Computing
Initiative. This work has made use of the resources provided by the 
Cambridge High Performance Computing service supported in part by 
the Science and Technology Facilities Council under grant ST/H008861/1.

\end{document}